\documentclass[12pt,letterpaper]{article}

\usepackage{amsmath}
\usepackage{amssymb}
\usepackage{graphicx}
\usepackage{pdfsync}
\usepackage{multirow}
\usepackage[numbers]{natbib}
\makeatletter 
\renewcommand\@biblabel[1]{#1.} 
\makeatother %
\usepackage{har2nat}
\usepackage{pdflscape}
\usepackage{setspace}
\usepackage{xcolor} 

\title{Transformation, normalization and batch effect in the analysis of mass spectrometry data for omics studies.
}
\author{Mertens, B. J. A.
}

\begin{document}
\maketitle\vspace{-0.5cm} \noindent Department of Medical Statistics, Leiden University Medical Center, PO Box 9600, 2300 RC
Leiden,  Netherlands
\\
\mbox{}
\\

\noindent {\bf Keywords}  \,\,\, Transformation, standardization, normalization, batch effect, scaling

\begin{abstract}
Data transformation, normalization and handling of batch effect are a key part of data analysis for almost all spectrometry-based omics data. This paper reviews and contrasts these three distinct aspects. We present a systematic overview of the key approaches and critically review some common procedures. Much of this paper is inspired by mass spectrometry based experimentation,  but most of our discussion carries over to omics data using distinct spectrometric approaches generally.
\end{abstract}

\setstretch{1.45}


\section{Spectrometry and data transformation}
There is a long-standing literature on the application of transformation in spectrometry data,  predating the advent of mass spectrometry. A good example may be found in the literature on infrared (IR) and near-infrared (NIR) spectrometry. An excellent recent introduction to statistical methods in this field was written by Naes et al \cite{Naes}.
The log-transform has a special significance in traditional spectrometry.
 An crucial component  of the appeal of the log-transform in (near) infrared spectrometry is due to Beer's law (1852) \cite{Beer1852},  which states that absorbance of light in a medium is logarithmically related to the concentration of the material through which the light must pass before reaching the detector. In other words,
\[
Absorbance=\log\Biggl(\frac{I_0}{I}\Biggr)=kLC,
\]
where $I_0$ is the incident intensity of the light and $I$ the measured intensity after passing through the medium, with $C$ the concentration, $L$ the path length the light travels through and $k$ the absorption constant. Another way to put this is that relative intensity is linearly related to concentration through the log-transform as
\[
\log\Biggl(\frac{1}{I}\Biggr)=\alpha+\beta C,
\]
where the path length, initial intensity and absorption constants are subsumed in the parameters $\alpha, \beta$.  Similar formulae exist for light reflection spectrometry. The formula has been used to provide justification for the application of classical linear regression procedures with log-transformed univariate spectrometric intensity readings.

\begin{figure}[!htbh]
\centering
\includegraphics[height=7cm]{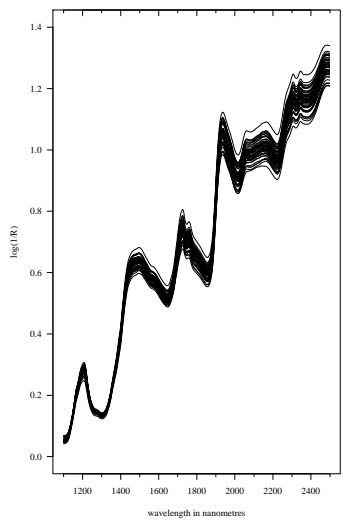}
\caption{\label{fig1}Log infrared reflectance measurements on 50 samples of mustard seeds within the 1100-2500 nanometer wavelength range. }
\end{figure}

The above constitutes an argument in favor of the log transform for spectrometry data based on non-linearity of spectral response.
It is partly responsible for causing the early literature on statistical and chemometric approaches in the analysis of spectrometry data to be based on the log-transformed measures.
 Beer's law  applies only to (univariate) IR or NIR spectrometric readings at a single wavelength. There is no multivariate extension of the law to cover full spectra consisting of the spectrometric readings across an entire wavelength range.
 Nevertheless, the log-transform would also be routinely applied once truly multivariate spectrometry became available,  jointly recording the spectrometric response at several wavelengths,  as shown in figure~\ref{fig1} which plots 50 NIR reflectance spectra across the range 1100-2500 nanometers.

Log transforming was heavily embedded  in the statistical spectrometry literature, when modern laser-based mass spectrometry measurement became routinely available by the end of the 20$^{th}$ century.
 Although Beer's law does  not apply to mass spectrometry,  the log-transform continues to be key to mass spectrometry based omics data analysis.   This is because the statistical reasons and rational behind the log-transform are more enduring and powerful  than the appeal to Beer's law might reveal.  We discuss 4 distinct arguments in favor of the log-transform.

\subsection{Scale and order of magnitude effects}
The first argument is associated with the objective of mass spectrometry to estimate protein composition in complex  mixtures consisting of  large numbers of proteins which may differ substantially in {\bf both} the {\it masses} of the constituent proteins in the mixture, but also in the {\it abundances} in which these are present.
It is precisely because of this objective that mass spectrometry has become a tool of choice in modern omics research.
Indeed modern mass spectrometric instruments are specifically engineered to have the capability to measure protein concentrations across large ranges of abundance, typically spanning several orders of magnitude -  known as the so-called dynamic range -  and to simultaneously achieve this across a wide mass range.

Unfortunately,  the spectrometry engineer's delight then becomes the statistical analyst's nightmare as this property renders data which spreads across several orders of magnitude in spectral response.
A good example may be found in Fourier-transform spectrometry,  which can easily display variation across 5 to 6 orders of magnitude.
In extreme cases this may cause numeric overflow problems in analyses,  although use of modern professional statistical software may reduce this problem.
Taking logarithms removes the order of magnitude effects.

\subsection{Skew and influential observations}
A second issue related to the above is that spectral measures will tend to be extremely skewed,  not only within an individual (across the within-sample spectrum responses), but also across samples or patients at a single m/z point. This may render the data unsuitable for standard analyses such as linear discriminant or similar when used on the original scale.

A related issue is that the skew may cause or be associated with a limited set of highly influential observations.
This is particularly troubling in omics applications,  as the spectra are typically very high-dimensional observations either when storing the response on a large grid of m/z values which can easily range in the thousands,  or after reduction to an integrated peak list.
Influential observations may affect the robustness of conclusions reported,  as results may differ substantially after removal of a single - or isolated group of - spectra from the analysis. A good example may be found in the calculation of a principal component decomposition as a dimension reduction prior to application of some subsequent data analytic procedure. Principal component analysis is known to be sensitive to extreme observations  \cite{Jolliffe2002}, particulary in high-dimensional applications with small sample size.  The same phenomenon will however also tend to apply for other analysis approaches,  such as regression methods,  discriminant procedures and so on. Transformation to log-scale may mitigate this problem.

\subsection{Statistical properties of particle counting}
At some risk of oversimplification,  mass spectrometers are in some sense nothing else but sophisticated particle counters, repeating a particle counting operation at each m/z position along the mass/charge range which is being investigated.  As a consequence,  spectrometry measures tend to have statistical properties reminiscent of those observed in Poisson (counting) processes. The variation of the spectral response tends to be related to the magnitude of the signal itself.
This implies that multiplicative noise models are often a more faithfully description of the data. Multiplicative error data are however more difficult to analyze using standard software. Log-transforming can be used to bring the data closer to the additive error scenario.

\subsection{Intrinsic standardization}
An interesting property of log transforming is that it may lead to intrinsic standardization of the spectral response.  Imagine for example that we are interested in calibrating the expected value $E(Y)$ of some outcome $Y$ based on two spectral responses $X_1$ and $X_2$ and that we have a study available recording both the observed outcome,  and the two spectrometry measures across a collection of samples (such as patients).  Let us also assume that we can use some generalized linear model to link the expected outcome to the spectrometry data via some link function $f$ and that the true model may be written in terms of a linear combination of the log-transformed spectral measures
\begin{align*}
f(E(Y))&=\alpha+\log(X_1)-\log(X_2)\\
\intertext{which reduces to }
f(E(Y))&=\alpha+\log \Biggl(\frac{X_1}{X_2}\Biggr).
\end{align*}
The result is that the linear dependence of the expected outcome via the link function on the log-transformed data actually implies regression on the log-ratio of the spectral responses $X_1$ and $X_2$, such that any multiplicative effect would cancel.  This can be regarded as an implicit form of standardization through  the log-transform. It is a general property which can be used with many statistical approaches such as (generalized) linear regression,  discriminant analysis and so on.

For spectrometry data generally and mass spectrometry particularly,  generally good advice would be to replace the raw measurements with log-transformed values at an early stage of the analysis, by application of the transformation
\[
\log(Y+a)
\]
with $a$  a suitably chosen constant.
Many statistical texts will also mention the Box-Cox transform when discussing the log transform.  For (mass) spectrometry data this approach is of limited value however,  because the optimal transform may lack the multiple justifications given above - which might as well be used as {\it a priori} grounds for choosing the logarithm -  but also because the approach is by definition univariate,  while a modern mass spectrometry reading will consist of thousands of measures across  m/z values and samples.

\section{Normalization and scaling}
Normalization is an issue which is often encountered in omics data analyses,  but is somewhat resistant to a precise definition.   We can identify what are usually perceived as the main objectives of it  and warn about the dangers associated with the topic,  so that we may avoid the most common pitfalls.

The objective of normalization can be loosely described as removing any unwanted variation in the spectrometric signal
which  cannot  be controlled for or removed in any other way, such as  by modifying the experiment for example.
This sets it apart from batch effect, which we will discuss later. The latter can sometimes  be adjusted for or accommodated by changing the experiment so that its effects can be either explicitly removed or adjusted for in subsequent analysis, by exploiting the structure of the experimental design. Not all effects can be accounted for in this manner however.

Examples of such effects which may induce a need for normalization are variations in the amount of material analyzed, such as ionization changes e.g.,  small changes in `spotting' sample material to plates, subtle fluctuations in temperature, small changes in sample preparation prior to measurement,  such as bead-based processing to extract protein, differential sample degradation, sensor degradation and so forth.  An important feature of such variation is that,  while we may speculate such variability sources are there and affect our experiment,  they are difficult to either control or predict,  which typically means all we can do is try to post-hoc adjust for it,  but prior to any subsequent analysis steps.

Important in devising an appropriate normalization strategy is that we should try to remove or reduce these effects on the measured spectral data,  while  retaining the relevant (biological) signal of interest. Unfortunately,  this is typically problematic for spectrometry.  This is because, as explained in the above paragraphs on transformation, the spectrometric signal and its variability are typically linked, often even after log-transform,  while the unwanted sources of variation affect all measures derived from the spectrum.

Imagine we have a study recording a mass spectrum on a dense grid of finely spaced points along the mass range or alternatively storing the data as a sequence of integrated peaks representing protein or peptide abundances  and this for a collection of samples (be it patients,  animal or other).  We write the ordered sequence of spectrometry measures for each $i^{th}$ sample unit as
${\bf x}_i=(x_{i1},\ldots,x_{ip})$,  with $p$ the number of grid points at which the spectrum is stored or the number of summary peaks. A transformation choice favored by some analysts is to apply early on in the analysis (possibly after first application of the log transform) standardization to unit standard deviation of all spectral measures at each grid point separately.  In other words, we replace the original data at each $j^{th}$ gridpoint with the measures $x_{ij}/std({\bf x}^j)$ where $std({\bf x}^j)$ is the standard deviation of the measures ${\bf x}^j=(x_{1j},\ldots,x_{nj})^T$ across all $n$ samples at that gridpoint. This procedure, sometimes also referred to as reduction to z-scores, is a form of scaling.  It is identical to standardization to unit standard deviation of predictor variables in regression analysis (\cite{Vach},  pages 124-125,  \cite{Snedecor}, page 349 and pages 357-358) when predictor variables are measured at different measurement scales (different units,  such as kg, cm, mg/l and so on). Indeed,  in the early days of regression analysis,  reporting standardized regression coefficients was an early attempt at assessing relative importance of effects.

\begin{figure}[!htbp]
\centering
\includegraphics[height=7cm]{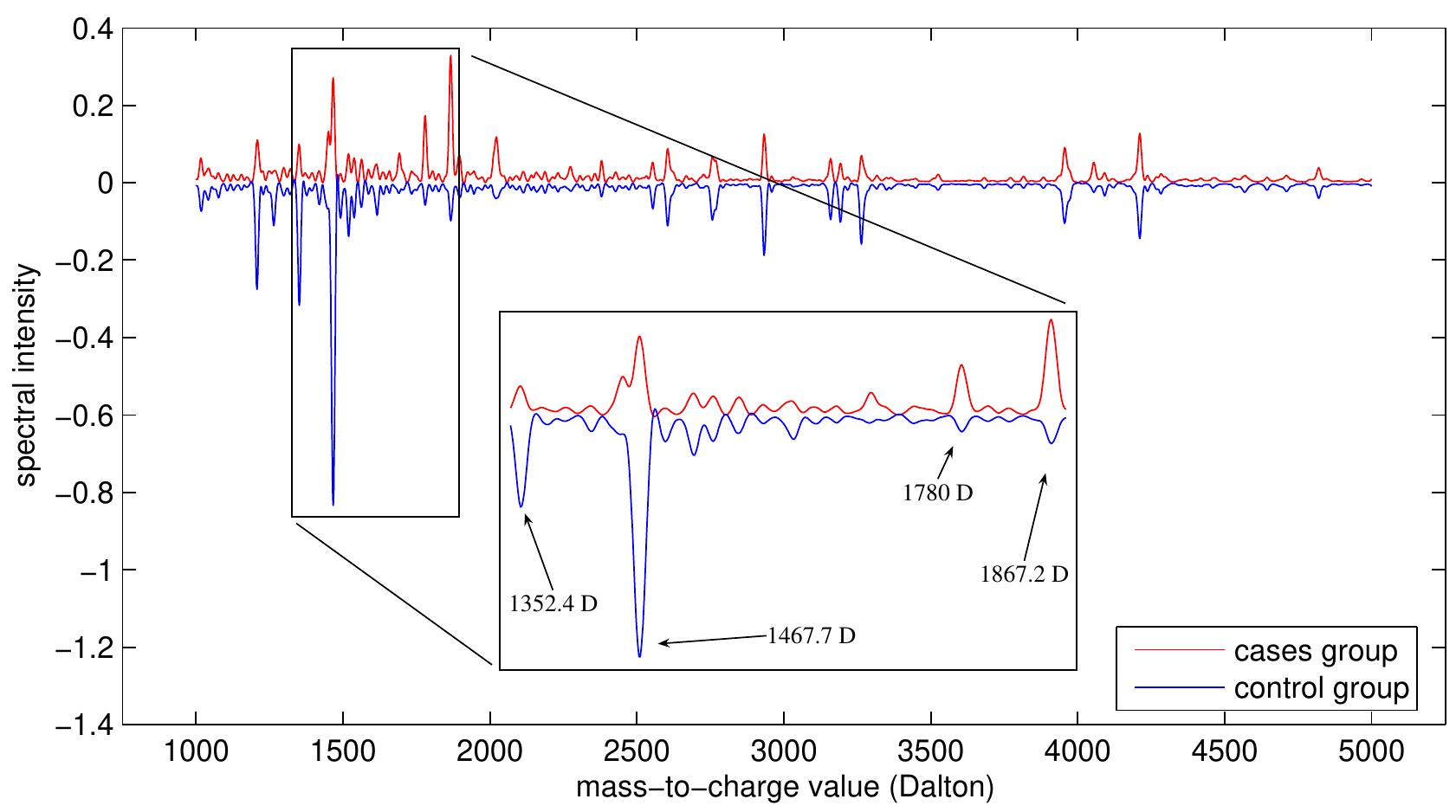}
\caption{\label{fig2}Spectra}
\end{figure}

\begin{figure}[!htbp]
\centering
\includegraphics[height=7cm]{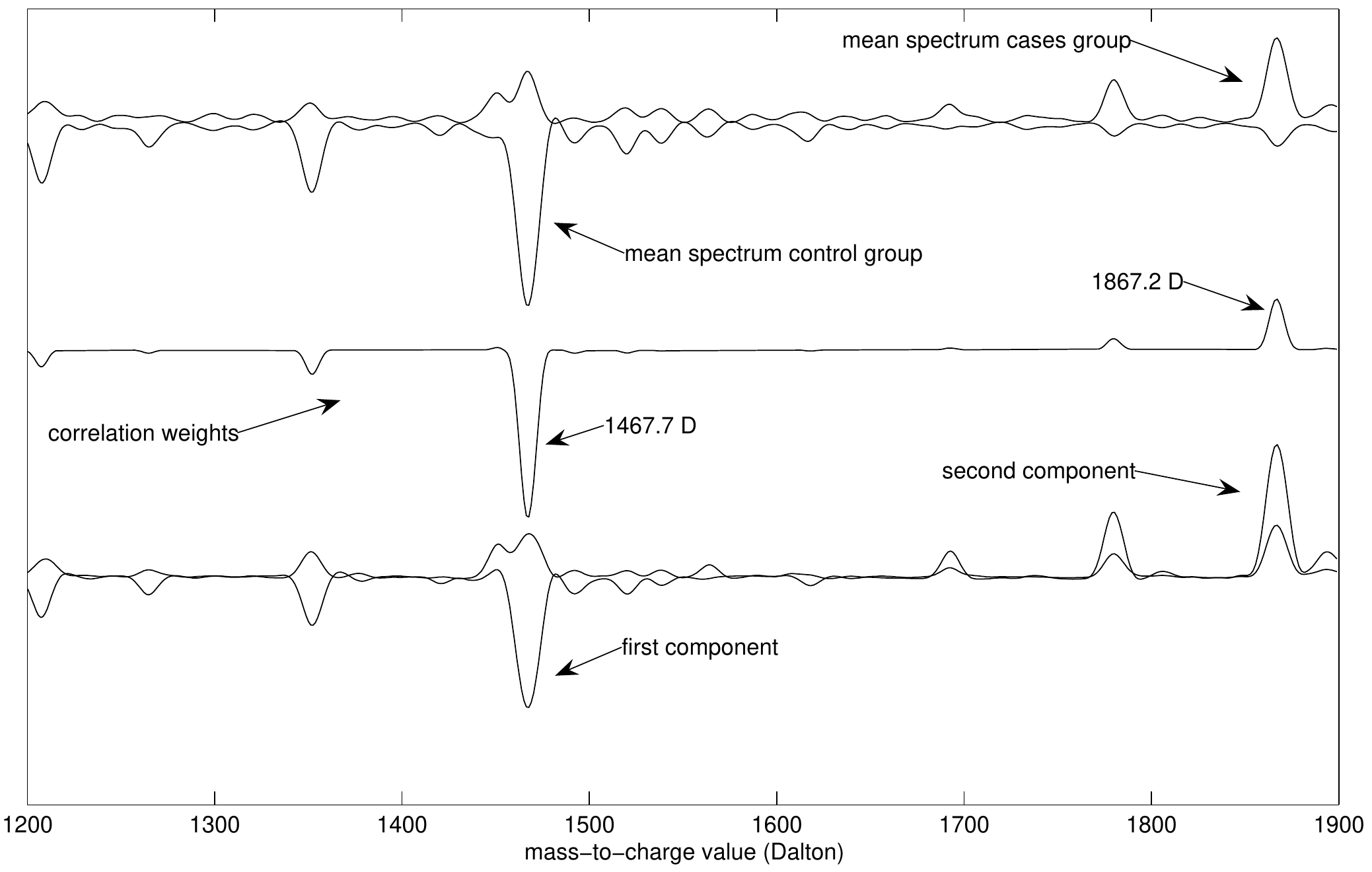}
\caption{\label{fig2_2}The top plot show mean cases and controls spectra separately.  The bottom curves are the loadings of the first two principal components across the same mass/charge range.  The middle curve shows the discriminant weights from a logistic regression model calibrated to distinguish cases from controls with the same data.}
\end{figure}

For spectrometry data generally and mass spectrometry in particular, standardization is more complicated.  Measurement units are by definition identical within a spectrum across the mass range. This would counsel against transforming to unit standard deviation as calibrated effects then remain directly comparable across the mass range on the original untransformed scale. There are however stronger arguments against this form of standardization in spectrometry.  Figure~\ref{fig2} illustrates the issues. The plot shows mean (MALDI-TOF) spectra from a clinical case-control study, after suitable transformation.
  To ease comparison, we plot the negative control spectrum versus the mean case spectrum. The rectangular region highlights and enlarges a region between 1200 and 1900 Dalton where most of the discriminant effects are found between the cases and control groups,  based on a discriminant analysis. Indicated are 4 key peaks at 1352.4, 1467.7, 1780 and 1867.2 Dalton which together summarize most of the between-group contrast between cases and controls. Figure~\ref{fig2_2} shows different statistics calculated on the same data within the same mass range.  The top plot again shows the mean spectra for cases and controls within the 1200-1900 Dalton region as before,  while the middle graph plots a graph of weighted discriminant coefficients obtained from a linear discriminant model calibrated from the data. It is obvious how the discriminant analysis identifies the peaks at 1354.2 and 1467.7 Dalton  and contrasts these with the peaks at 1780 and 1867.2 Dalton.  The below graph in figure~\ref{fig2_2} shows the first two principal components calculated on the same data and based on the pooled variance-covariance matrix.

There are several things to note in this picture.  The first is how much the principal component and mean spectra curves resemble one another. The first component closely approximates the mean control spectrum,  while the second component does the same for the mean cases spectrum.
At first sight,  this might seem all the more remarkable, since the principal component decomposition is based on the pooled variance-covariance matrix,  and hence on the `residual spectra' ${\bf x}_i-\overline{\bf x}_{g(i)}$, where $\overline{\bf x}_{g(i)}$ denotes the mean spectrum of group $g(i)$ to which the $i^{th}$ observation belongs, with $g=1,2$ for the cases and control groups respectively. So the figure shows two different aspects of the data. One is the systematic (mean) spectral response (top graphs),  the other are the deviations relative to the mean spectral outcome (bottom graphs). From the figure,  we can see that the component decomposition tells us that the peaks at 1352.4 and 1467.7 Dalton are highly correlated and account for much of the variation in the spectral data,  as they weigh heavily in the first principal component. Similarly,  the second component summarizes much of the expression in both peaks at 1780 and 1867.2,  which are again highly correlated. Because of this,  the classification might as well be summarized as a contrasting between the first and second principal component, since this would contrast peaks 1352.4 and 1467.7 with the expression at peaks 1780 and 1867.2 (see Mertens et al \cite{Mertens2006} for the full analysis).

This feature of the data where the mean expression and deviations from the mean are closely linked as shown in the above example,  is typical of spectrometric variation.  It is the consequence of the connection between mean expression and variance we mentioned above when discussing the log-transform and can be observed in almost all spectrometry data,  often even after log transforming. To put this differently,  in spectrometry data,  we will find the signal where the variance is (even if we correct the variance calculation for systematic differences in expression,  as shown in our above example).  It is for this reason that transforming to unit standardization should be avoided with spectrometry data, unless scale-invariant methods are explicitly used to counter this problem.

In addition to the above considerations,  there are also other arguments for avoiding reduction to z-scores or transformation to unit standard deviation.  An important argument here is that summary measures such as means and standard deviations are prone to outliers or influential observations,  which can be a particular problem in high-dimensional statistics and with spectrometry in particular. A specific problem with such form of standardization is that it may cause problems when comparing results between studies. This is because systematic differences may be introduced between studies (or similarly, when executing separate analyses between batches - see further), due to distinct outliers which affect the estimates of the standard deviations for standardizing between repetitions of the experiment.

Our final comments on the above standardization approach is that medians and inter quartile ranges (IQR) are sometimes used instead of means and standard deviations in an attempt to alleviate some of the robustness concerns.  Other authors advocate use of some function of the standard deviation,  such as the square root of the standard deviation instead of the standard deviation itself.  This is sometimes referred to as  Pareto scaling \cite{Berg}.  The rational for this amendment is that it upweights the median expressed features without excessively inflating the (spectral) baselines. An advantage of the approach may be that it does not completely remove the scale information from the data.  Nevertheless, the choice of the square root would still appear to be an ad hoc decision in any practical data analytic setting.

\begin{figure}[!htbp]
\centering
\includegraphics[height=5cm]{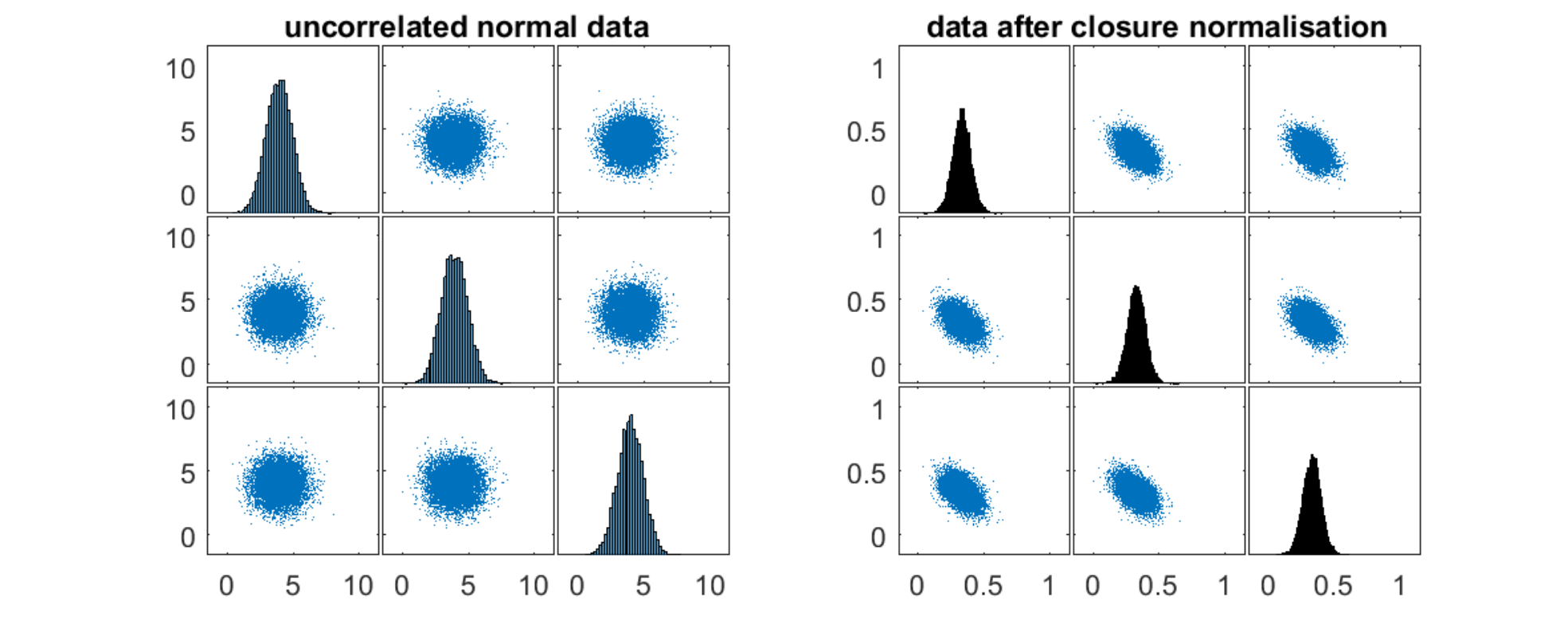}
\caption{\label{fig4}Effect of closure normalization on uncorrelated data.}
\end{figure}

Some authors make a formal distinction between scaling and normalization methods and consider the first as operations on each feature across the available samples in the study \cite{Craig}. Normalization is then specifically defined as manipulation of the observed spectral data  measurements on the same sample unit (or collection of samples taken from the same individual)(within-spectrum or within-unit normalization).
A potential issue with the above described approaches to normalization via statistical transformation is that they are based on a borrowing of information across samples within an experiment.
Another  extreme form of such borrowing is a normalization approach which replaces the original set of spectral expression measures for a specific sample with the sample spectrum measures divided by the sample sum,  such that the transformed
set of measures adds to 1. An argument sometimes used in favor of such transformation is that it would account for systematic differences in abundance - possibly caused by varying degrees of ionization or similar effects from sample to sample -  such that only the relative abundances within a sample are interpretable. Although this approach is unfortunately common,  it has in fact no biological foundation \cite{Craig}. Even if arguments based on either the physics or chemical properties of the measurement methodology could be found, these could not be used  in favor of  such data-analytic approach as described above,  which we shall refer to as `closure normalization'. The problem with the approach is that it actually {\it induces} spurious - and large - {\it biases} in the correlations between the spectral measures which mask the true population associations between the compounds we wish to investigate.
Figure~\ref{fig4} shows the effect of closure normalization on uncorrelated normal data in 3 dimensions. The left plot shows scatterplots between each of the
three normally distributed measures. The right shows scatterplots of the resulting transformed variables after closure normalization. The absolute correlations between each variable pair has increased from 0 (for the original uncorrelated data) to 0.5 (after transformation).
This becomes particularly problematic should the subsequent objective be to perform some form of network or association analysis, in which case the prior closure normalization renders results meaningless.  Similarly problems would however also apply to regression and discriminant  analysis.

A variant of the above closure approach to normalization which is sometimes also used is to adjust to the maximum peak observed in a spectrum,  where the maximum is either the spectrum-specific maximum, or the maximum at that spectral location which corresponds to the maximum mean spectrum across several samples. Adjustment is often carried out by dividing each spectrum by its maximum at the maximum location,  such that the spectral response gets a constant expression at the maximum location in the transformed data. Just as for closure normalization,  this procedure appears appealing on intuitive grounds at first sight,  but suffers from similar problems,  as the variation at the maximum induces severe correlations across the spectral range in the transformed data,  which {\it cannot have biological interpretation}. Both approaches should be avoided.

Data normalized by the sum of the combined expression (closure normalization) can be viewed as an instance of compositional data \cite{Aitchison}. Hence, instead of applying such normalization, one could therefor think of using special-purpose methods from the compositional data analysis literature, or to develop or adapt such methods for application in omics applications. This has not been attempted to our knowledge at time of writing.  As an alternative, it should be recommended to take a conservative approach and refrain from excessive transformation when the consequences are not well understood or accounted for in subsequent analysis.  In such cases restricting to log-transformation as discussed earlier is safer. In any case, the original untransformed data should always be at hand and stored to allow verification of results through possible sensitivity analysis.

The above is only an introduction to some of the main forms of normalization in use at this time.  Many other forms exist and will undoubtedly continue to emerge. An interesting one worth mentioning is the idea of `lagging' the spectra by taking differences between subsequent values within the spectral range.  With log-transformed spectrometric data, this is another approach which induces ratios between subsequent spectral intensities which eliminate multiplicative change effects. An example is found in an interesting paper by Krzanowski et al \cite{Krzanowski}. It has the drawback that results from subsequent statistical analysis can be more difficulty to interpret,  but it might be of use in pure prediction problems. Other forms of standardization and normalization are also found in the literature, particularly methods which seems to inherit more from common approaches in microarray analysis,  such as quantile normalization \cite{Eidhammer}. Ranking of spectral response, including the extreme form of reducing to binary have also been investigated.  The latter can be particularly useful as a simple approach when data are subject to a lower detection limit \cite{Kakourou}. Other forms of normalization and standardization worth mentioning at time of writing are scatter correction and orthogonal signal correction.  We refer to Naes et al \cite{Naes} for a good introduction to these methods.

Which transformations should be applied first?  What is a good order of applying distinct normalization or transformation steps? There is some difference of opinion between researchers on the precise sequence in which various normalization procedures are applied to the data.  As a general rule it seems wise to apply logarithms early and calculate means and standard deviations only after log-transforming.

The issue of normalization is closely linked to the problem of standardization of mass spectra. Several definitions of standardization may be possible here. One option is to define the problem as  `external' standardization, which would form part of the experiment itself (as opposed to the post-experimental data processing we describe before) where we somehow try to change the experiment so that part of the systematic experimental variation is either prevented from occurring or could be accounted for through post-processing of the data.  Examples would be in the use of spike-in controls,  on a sample plate,  or even within the sample material itself,  so that the spectral response can  be adjusted for the expression of the known spike-in material which is added.  Another example would be in the use of technical controls on a sample plate with know concentrations.  Yet another example would be systematic equipment re-calibration to re-produce a (set of) known standards,  so that sample-to-sample variation due to experimental drift is suppressed as much as possible. All these approaches to standardization are different from the above described methods in that they try to circumvent known sources of variation by changing the experiment itself,  rather than post-hoc attempting to adjust for it.

\section{Batch effect in omics studies}
A batch effect is a source of unwanted variation within an experiment which is typically characterized by the property that it is due to a known structure within the experiment.
Usually the structure, and thus also the existence of the associated effect, is known in advance of the experiment but cannot be avoided or eliminated from the measurement process.
Furthermore, the effect of the experimental structure itself may not necessarily be predictable either,  even when knowledge of the structure exists.
A typical example would be distinct target plates which are used to collect sample materials prior to analysis in some mass analyzer.
However, we can sometimes manipulate or modify the experiment  to account for the known structure so that it is rendered innocuous and  cannot unduly affect conclusions.
Even better,  we may be able to remove (part of) the effects due to batch structure, by taking suitable precautions at the experimental design stage.

In contrast to genetics,  batch effect is much more problematic within proteomics and metabolomics,  which is due to the measurement procedures involved.
It should also be noted that batch effect and the accommodation of it are different from standardization and normalization issues in that we need to identify its presence and consider its potential effects both {\it before} and {\it after} the experiment.
Before the experiment,  because we may want to tweak or change the experimental structure to take the presence of the batch effect into account (this may involve discussion between both statistician and spectrometrist).
The objective is to change the experimental design so as to avoid confounding of the batch structure with the effect or group structure of interest.
After the experiment,  because we may wish to apply some data-analytic approach to remove the effect (which may depend on the experiment having been properly designed in the first place).
Accounting for batch effect in proteomic studies will hence involve two key steps.
\begin{enumerate}
\item  To control the experimental design as much as possible in advance of the experiment to maximize our options to either remove or account for batch effect at the analysis stage.
\item  To exploit the chosen structure afterwards  to either remove or adjust for the batch variation in the analysis.
\end{enumerate}
The objectives of these steps are at least threefold.
\begin{enumerate}
\item To ensure experimental validity (the experiment can deliver the required results)
\item To improve robustness of the experiment (conclusions will still be valid,  even when experimental execution differs from experimental planning)
\item To improve (statistical) efficiency of effect estimates based on the data (statistical summaries will have lower variation)
\end{enumerate}
In the following discussion on batch effect we will make a distinction between the following two distinct types of batch effect which may occur in practical experimentation.
The first are {\bf time-fixed batch effects}. Examples of these are
\begin{itemize}
\item plates in mass spectrometry
\item freezer used for sample storage
\item change of cleaning or work-up fluid for sample processing
\item batches of beads or cartridges for protein fractionation
\item instrument re-calibration
\end{itemize}
In contrast, {\bf time-dependent batch effects} are due to experimental structures associated with time. Examples of these are found in the following situations.
\begin{itemize}
\item longitudinal experimentation, long observation tracks with repeated measurement per individual
\item sample collection across extended time periods, patient accrual spread across an extended period of time
\item time-indexed samples or instrument calibrations
\item distinct days of measurement or sample processing
\end{itemize}
Both types of batch effects may occur in proteomic experiments,  but for different reasons and with distinct consequences.  The treatment of both types of batch effect will also be different between the two.

\subsection{Time-fixed batch effects}
\begin{figure}[!htbp]
\centering
\includegraphics[height=7cm]{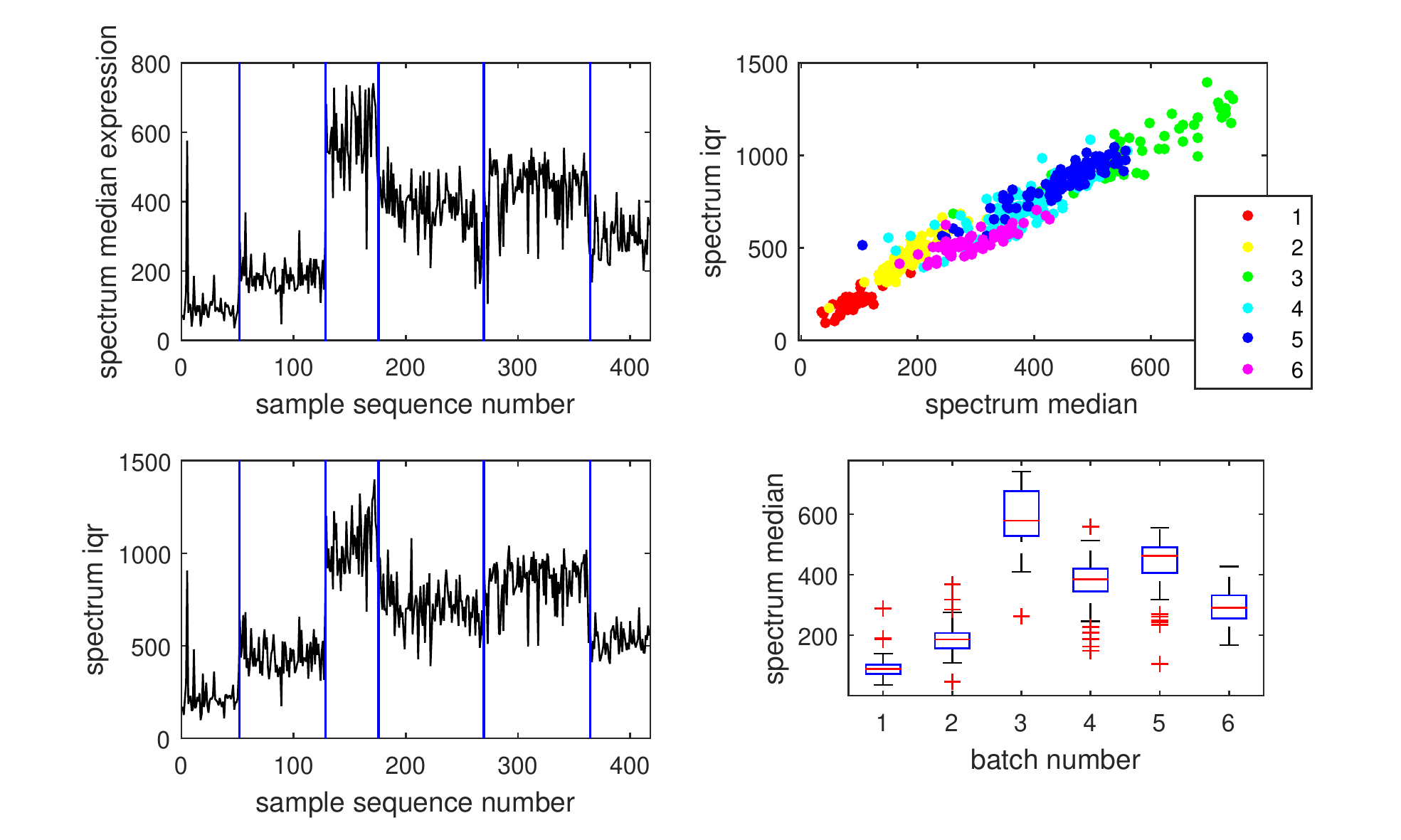}
\caption{\label{fig3}Batch effect in a case-control study using 6 target plates. The vertical lines in the left-side plots indicate transition to a new plate (batch). The distinct colors in the scatter plot represent the different batches.}
\end{figure}

We consider a case-control study as an example.
 The experiment contrasted 175 cases with 242 healthy controls. Due to the large sample size and multiple replicate measurements per sample, sample material needed to be assigned to six target plates prior to mass spectrometric measurement.
The plates constitute a systematic - batch - structure within the experiment. On inspecting within-sample medians and inter-quartile ranges (IQR),  systematic plate-to-plate differences were noted, shown in figure~\ref{fig3}.
Analysis of the data using a discriminant approach led to correct classification of 97\% of samples. Unfortunately, on closer investigation of the experimental design,  it turned out that all case material had been assigned sequentially to the first 3 plates,  after which the controls samples were thawed next and assigned to the subsequent plates. It is not possible to statistically adjust for such perfect confounding between the plate structure and the potential between-group effect which was the primary target of the research study. After discussion, the study was abandoned, leading to significant loss of both experimentation time and resources,  among which the valuable sample materials.

A simple procedure exists to prevent such problems, called blocked randomization. It consists of assigning cases and controls in  equal proportions and at random to the distinct plates. Table~\ref{Tab1}
shows such a design for a case-control study randomizing cancer cases and healthy controls to three target plates, as reported by Mertens et. al. \cite{Mertens2006}, which gives more details about the study. In addition to randomizing the cases and controls in roughly equal proportions across the plates,  the study also tried to have cancer stages in roughly equal distributions from plate-to-plate. A recent overview of classical principles of statistical design for proteomics research can be found in a recent paper by Cairns \cite{Cairns} and the references therein.

\begin{table}[!htbp]
{\scriptsize
\begin{center}
\begin{tabular}{l  cccc cc cccc cc cccc c}
\hline
Group&\multicolumn{16}{l}{Plates}& Total\\
&1& &&& &&2 & &&& &&3\\
\\

Controls           &17 &&&    & &&17  &&&  &  &&16 &&& & 50\\
Cases              &22 &&&    & &&22  &&&  &  &&19 &&&  & 63\\
\\
\multicolumn{1}{r}{Stage}&\multicolumn{1}{l}{1}&\multicolumn{1}{l}{2}&\multicolumn{1}{l}{3}&\multicolumn{1}{l}{4}&&
&\multicolumn{1}{l}{1}&\multicolumn{1}{l}{2}&\multicolumn{1}{l}{3}&\multicolumn{1}{l}{4}&&
&\multicolumn{1}{l}{1}&\multicolumn{1}{l}{2}&\multicolumn{1}{l}{3}&\multicolumn{1}{l}{4}                        \\
\multicolumn{1}{r}{Cases}&\multicolumn{1}{l}{4}&\multicolumn{1}{l}{10}&\multicolumn{1}{l}{4}&\multicolumn{1}{l}{4}&&
&\multicolumn{1}{l}{4}&\multicolumn{1}{l}{10}&\multicolumn{1}{l}{4}&\multicolumn{1}{l}{4}&&
&\multicolumn{1}{l}{3}&\multicolumn{1}{l}{8}&\multicolumn{1}{l}{4}&\multicolumn{1}{l}{4}
\\
\hline
\end{tabular}
\end{center}
\caption{\label{Tab1}Block-randomized case-control study in a study using 3 plates. The design is shown \underline{as executed}
on the first week. A replicate of the entire experiment was run on
the subsequent week using plate duplicates. `Stage' refers to the
distribution of cases across the four respective disease stages.}
}
\end{table}

Another example is a glycomics study which assigned cases and controls to target plates sequentially as the samples  became available.
By the end of the study period, 288 samples had become available,  of which 97 were cases and 191 controls. The case-control assignment to plates is shown in table~\ref{Tab2}. As can be seen, case-control assignment is perfectly confounded with the plate effect for 25 out of 34 experimental batches,  which makes these measurements useless for between-group comparison. Note also how only two plates, indicated in red typescript contain appreciable numbers of {\it both} cases and controls.
\begin{table}[!htbp]
\begin{small}
\begin{center}
\begin{tabular}{|r|r|r|r|r|r|r|r|r|r|r|}
\cline{1-2} \cline{4-5} \cline{7-8} \cline{10-11}
ca&co&&ca&co&&ca&co&&ca&co\\
\cline{1-2} \cline{4-5} \cline{7-8} \cline{10-11}
4&0&&1&3&&0&15&&1&6\\
\cline{1-2} \cline{4-5} \cline{7-8} \cline{10-11}
3&0&&0&9&&1&0&&1&9\\
\cline{1-2} \cline{4-5} \cline{7-8} \cline{10-11}
11&0&&4&0&&3&0&&0&7\\
\cline{1-2} \cline{4-5} \cline{7-8} \cline{10-11}
5&0&&1&0&&0&4&&0&4\\
\cline{1-2} \cline{4-5} \cline{7-8} \cline{10-11}
12&0&&1&0&&1&0&&&\\
\cline{1-2} \cline{4-5} \cline{7-8} \cline{10-11}
\textcolor[rgb]{1,0,0}{21}&\textcolor[rgb]{1,0,0}{40}&&1&0&&0&5&&&\\
\cline{1-2} \cline{4-5} \cline{7-8} \cline{10-11}
1&3&&2&0&&1&9&&&\\
\cline{1-2} \cline{4-5} \cline{7-8} \cline{10-11}
2&0&&0&4&&2&8&&&\\
\cline{1-2} \cline{4-5} \cline{7-8} \cline{10-11}
\textcolor[rgb]{1,0,0}{16}&\textcolor[rgb]{1,0,0}{13}&&0&3&&0&4&&&\\
\cline{1-2} \cline{4-5} \cline{7-8} \cline{10-11}
0&15&&0&16&&2&14&&&\\
\cline{1-2} \cline{4-5} \cline{7-8} \cline{10-11}
\end{tabular}
\end{center}
\end{small}
\caption{\label{Tab2}Case-control study assigning samples to 34 plates as they became available.  The assignment order is from top-to-bottom,  left-to-right.}
\end{table}
After analysis of the data,  it was found that the estimate of the batch effect (std) substantially exceeded the measurement error estimate (std). No clear evidence of differential expression of glycans between groups emerged,  though it could be hypothesized that any differences might be small and at least smaller than the observed between-batch variation. This raised questions whether the `design' was to blame for the failure of the study to identify differential expression.

To investigate the consequences of using a `design' as used in the above glycomics study,  we investigate a simulation study which contrasts several potential alternative designs.  For each of these 4 designs we assume the same sample size of 288 samples with 97 cases and 191 controls,  just as for the original study. We consider the following alternative scenarios.
\begin{enumerate}
\item Experiment 1 is a hypothetical experiment which is able to use only one plate (no plate effects present).
\item Experiment 2 is a randomized blocked version of the original experiment which uses 3 plates and distributes the cases and controls in approximately balanced ratios of 32/65, 32/63 and 33/63 to the first, second and third plate respectively.
\item Experiment 3 is a perfectly confounded experiment which again uses 3 plates,  but assigns all 97 cases to the first plate and assigns 95 and 96 controls to the other two plates respectively.
\item The last experiment uses the original glycomics experiment distribution of cases and controls to 34 plates, as  shown above in table~\ref{Tab2}.
\end{enumerate}
We now simulate experimental data for each of the above 4 scenarios, generating effect sizes ranging from 0 to 1.5 for a {\it single} glycan (univariate simulation) and assuming between-batch effects with standard deviations $\sigma_B$ taking values 3.6, 1.8, 0.9 and 0.45. The standard deviation of the error $\sigma_E$ takes the value 1.8 throughout (these numbers inspired by results from the real data analysis).

\begin{figure}[!htbp]
\centering
\includegraphics[height=6.5cm]{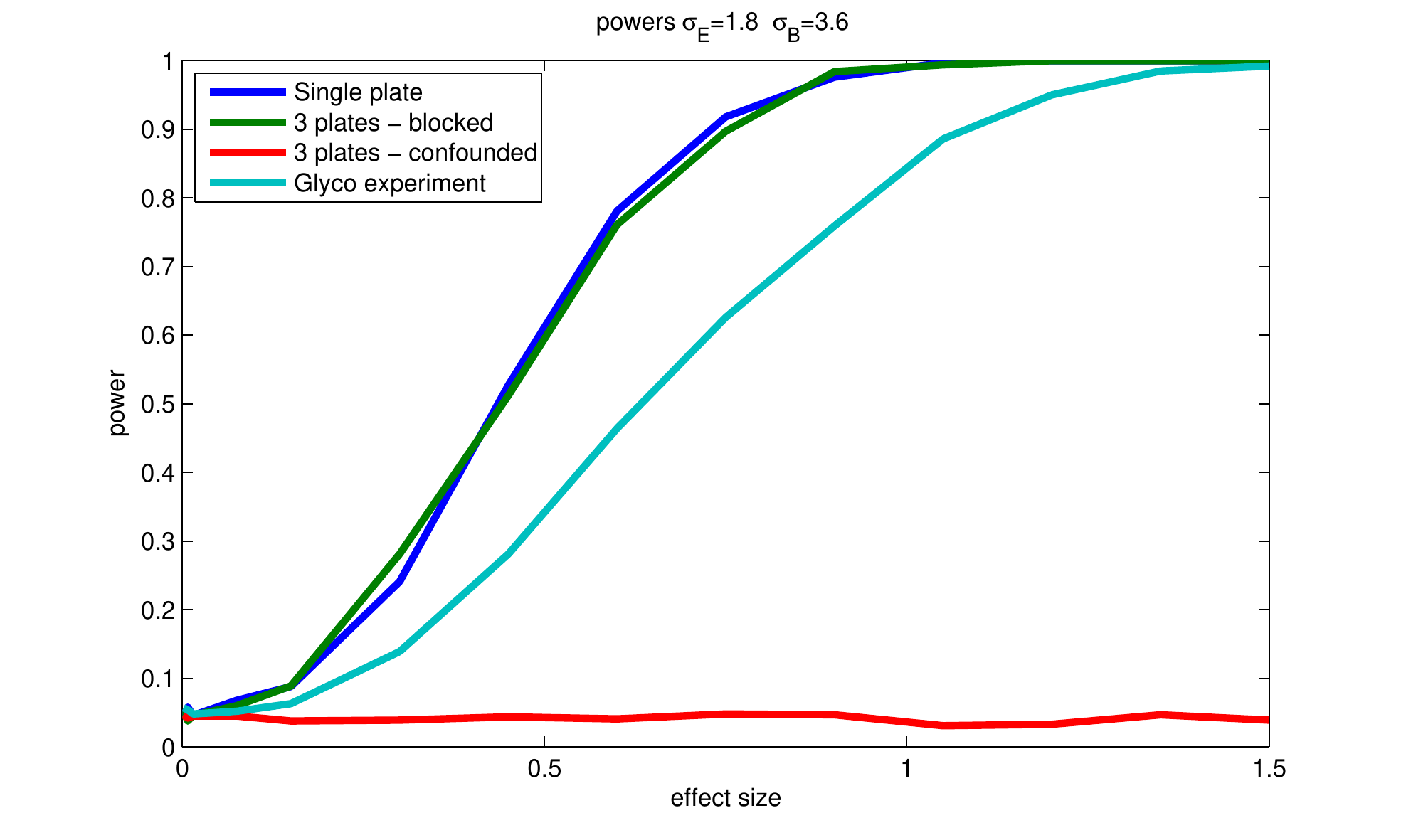}
\caption{\label{fig10}Spectra}
\end{figure}

\begin{figure}[!htbp]
\centering
\includegraphics[height=6.5cm]{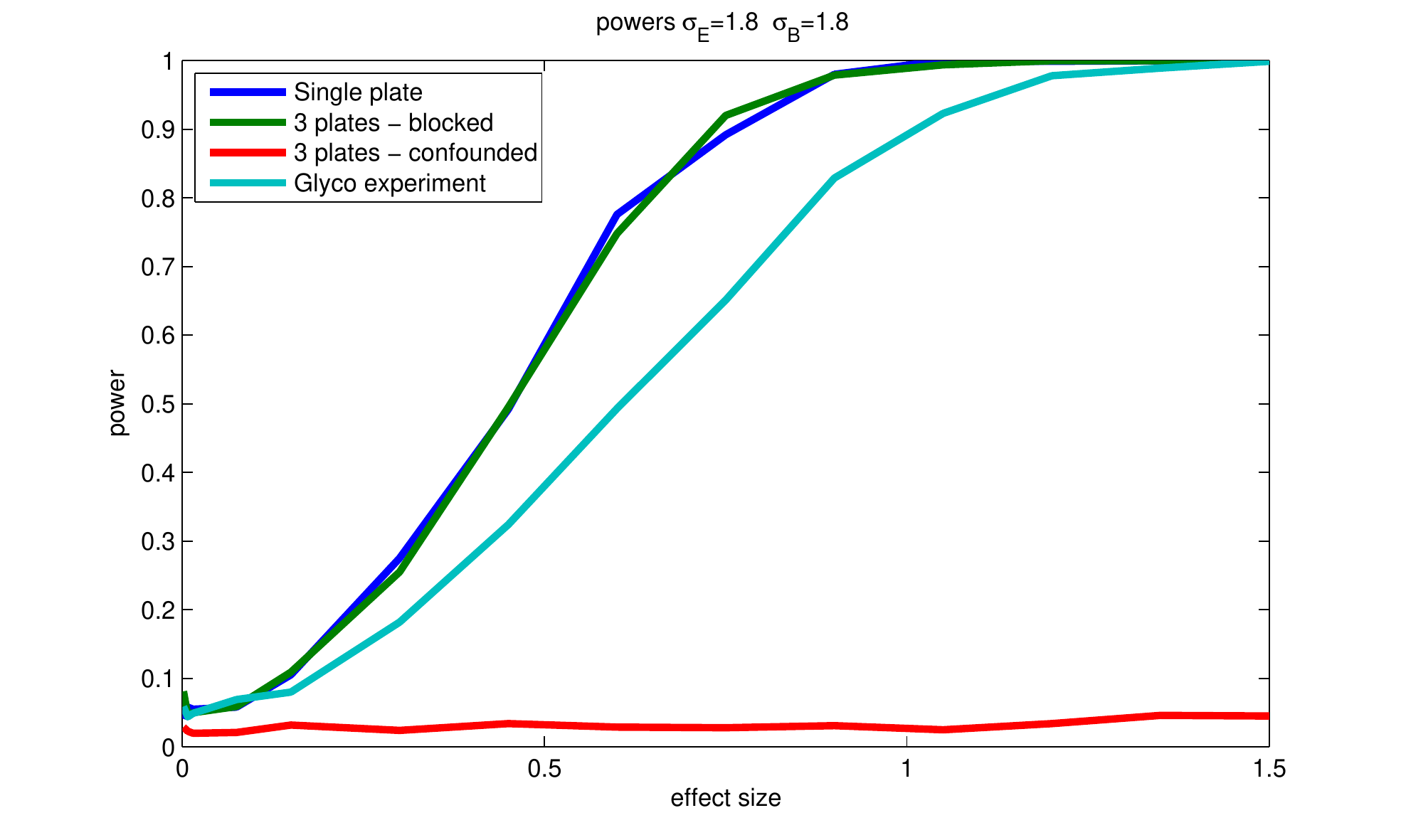}
\caption{\label{fig11}Spectra}
\end{figure}

\begin{figure}[!htbp]
\centering
\includegraphics[height=6.5cm]{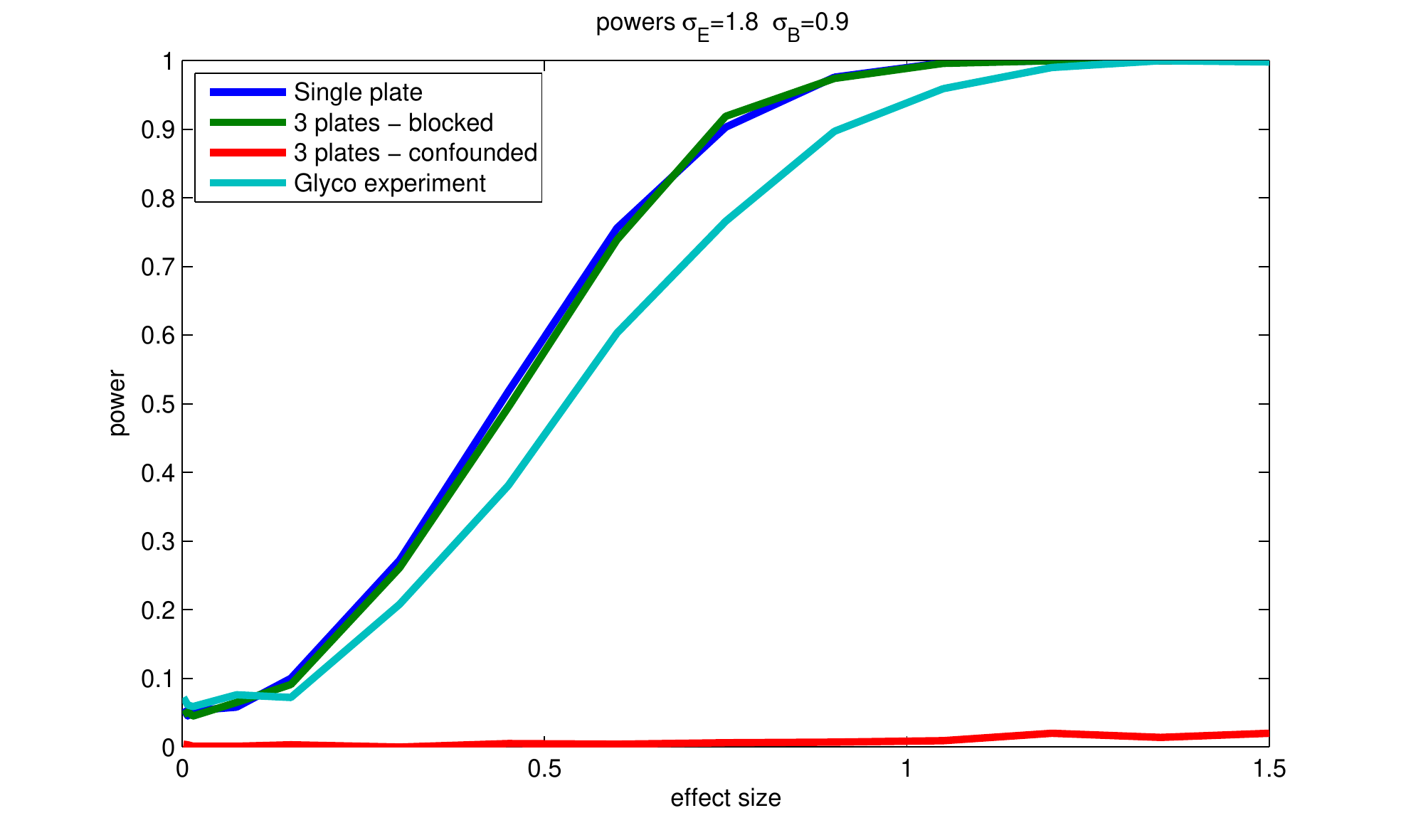}
\caption{\label{fig12}Spectra}
\end{figure}

\begin{figure}[!htbp]
\centering
\includegraphics[height=6.5cm]{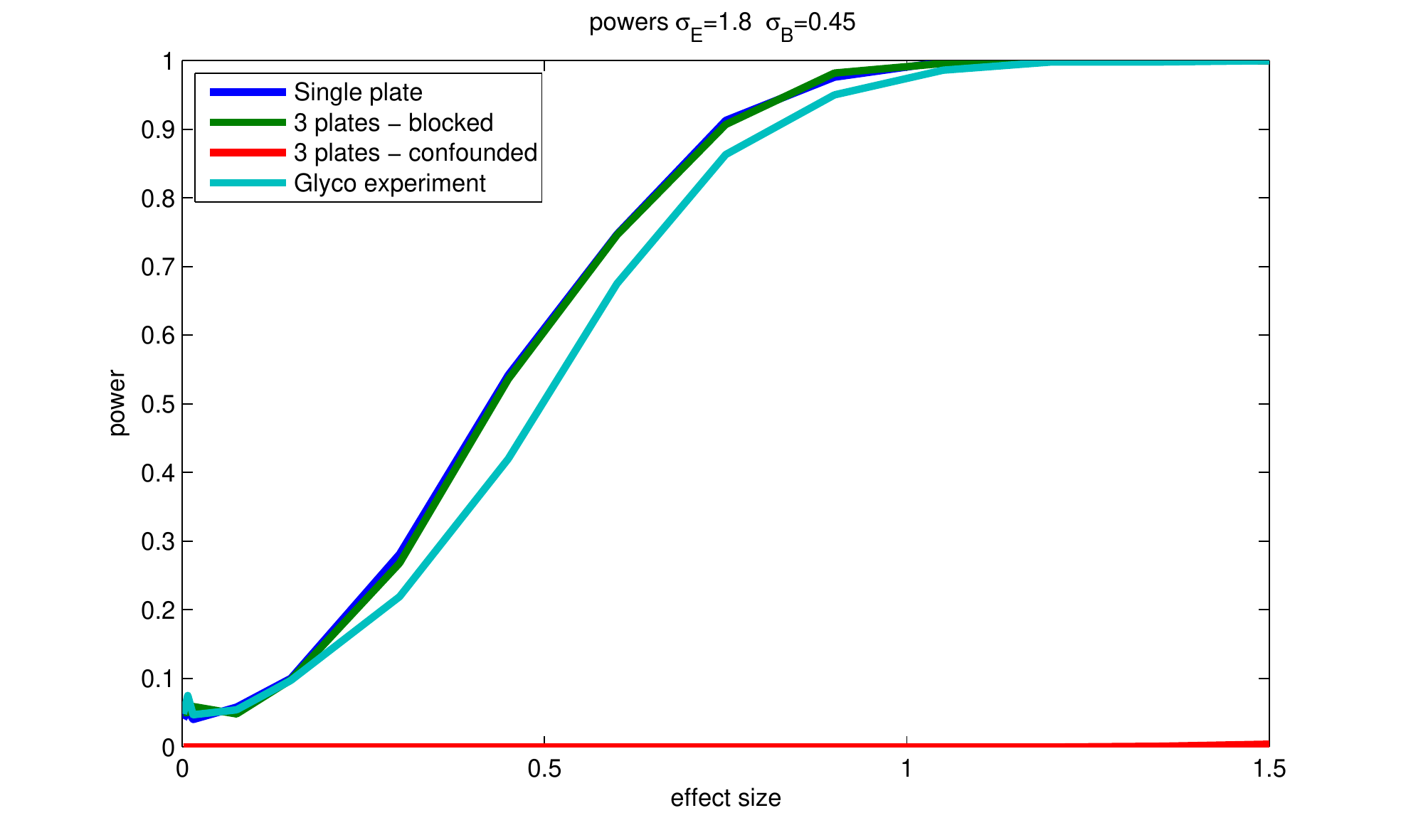}
\caption{\label{fig13}Spectra}
\end{figure}

\begin{figure}[!htbp]
\centering
\includegraphics[height=6.5cm]{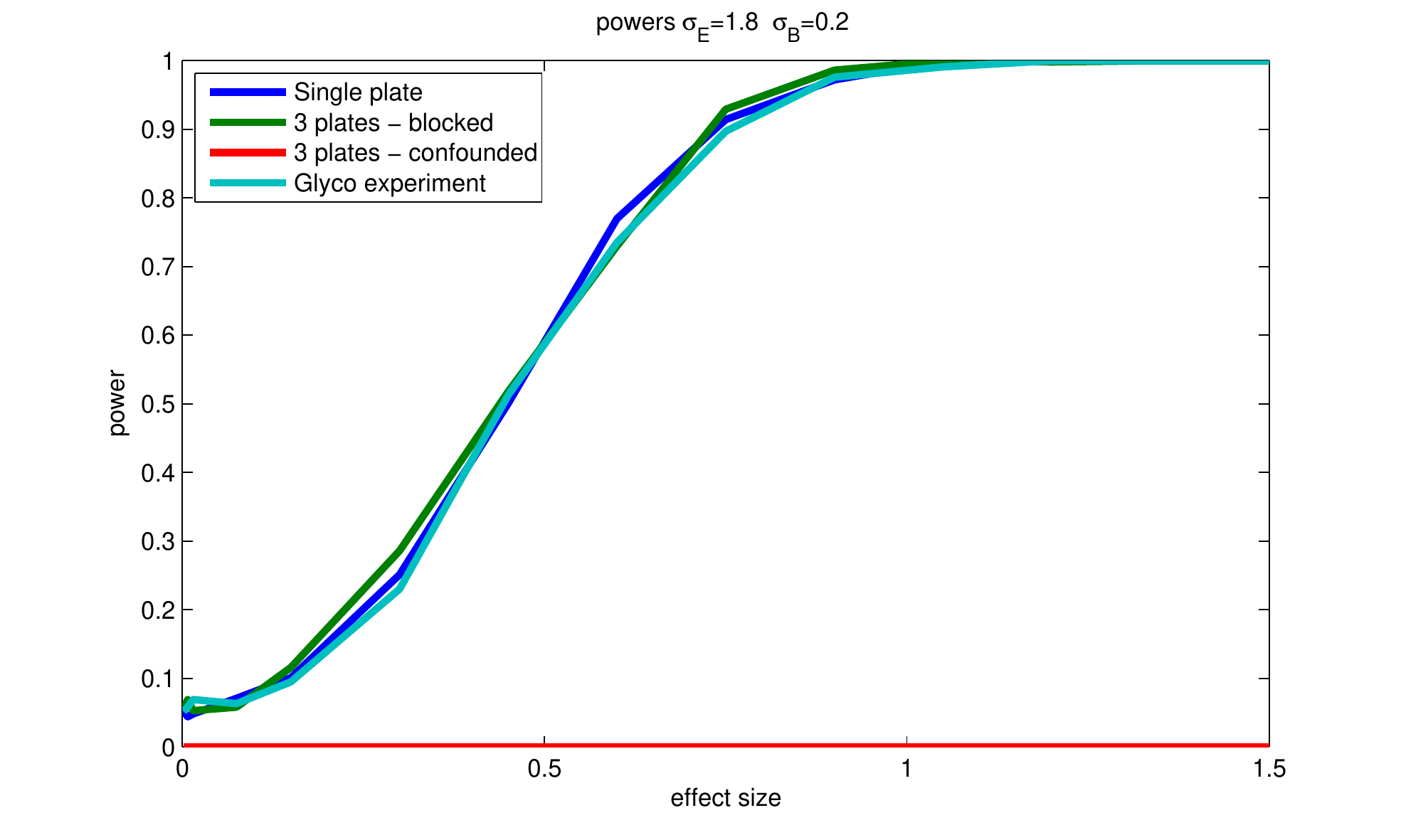}
\caption{\label{fig14}Spectra}
\end{figure}

In the analysis of the simulated data for the above experiments,  we fit linear mixed effect models \cite{Molenberghs} to the simulated data of experiments 2, 3 and 4. The mixed effect models correct for the known batch structure using a random effect while estimating the between-group effect with a fixed effect term.
For the first simulated experiment a simple linear regression model is used, which is equivalent to a two-sample pooled t-testing approach.
Figures~\ref{fig10} to \ref{fig14} shows the probabilities to detect the between-group effect (power) across the effect size range simulated and for the standard deviations of the batch effect indicated. As expected, we find that the probability to detect the effect increases as the effect size grows for the single-plate experiment [1]. The blocked experiment [3] matches the power of the single-plate experiment regardless of the size of the batch random effect. The power of the actually implemented glycomics experiment [4] depends on the size of the batch effect.  As the batch effect gets smaller (associated std goes to zero) then we can eventually ignore the batch effect altogether and we obtain the same powers as if the batch effect was not present. This is of course a confirmation of what we would expect to find. If the batch effect is substantial however (associated std is large relative to the size of the between-group effect),  then we pay a penalty in terms of seriously reduced powers of detecting the between-group effect. The perfectly confounded experiment [3] performs dramatically whatever the batch effect is,  since we are forced to account for a known batch structure - irrespective of the true but unknown population batch effect - and thus loose all power of detecting the effect of interest. The excellent performance of the blocked version of the experiment again emphasizes the need and importance of pro-actively designing and implementing block-randomized experiments when  batch structures are identified in advance of the experimentation.

\subsection{Time-dependent batch effects}
\begin{figure}[!htbp]
\centering
\includegraphics[height=6.5cm]{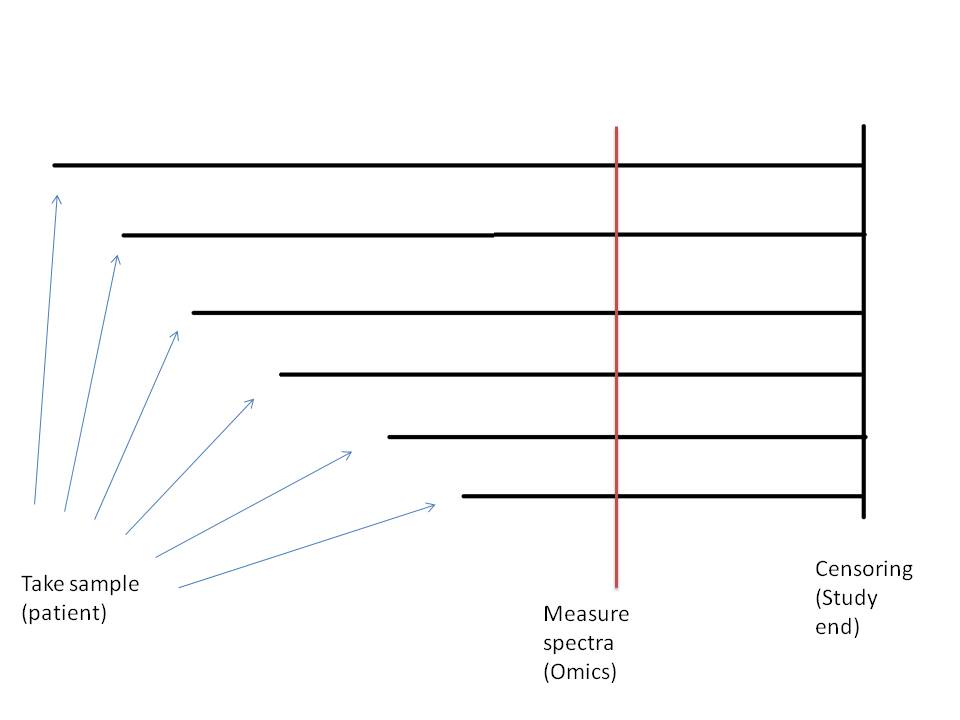}
\caption{\label{fig20}Spectra}
\end{figure}

Longitudinal experimentation with spectrometry-based omics is still in its infancy,  but we should expect this area to mature.
Experience with both analysis of such data,  but also pre-processing aspects and the handling of batch effects, must still develop.
We can however already point to some aspects which are likely to become a key concern in future applications for which omic science and technology will have to provide credible answers.
A key aspect which is liable to be an issue in many omic studies is illustrated in figure~\ref{fig20}.
The figure displays a study design where samples are collected dispersed over time and kept in storage until a decision is taken to extract sample material and analyze the samples using some form of spectrometry at a common point in time.  Patients themselves are followed up until end of the study - up to which point some outcome measure may be continually recorded during the follow-up - or some pre-defined endpoint (such as cancer occurrence or recurrence or similar outcome) prior to the study termination.
A common and relatively simple form of such a design which is often used in clinical research could for example be a survival study \cite{Cox}\cite{Klein}
with so-called `administrative censoring' as depicted in figure~\ref{fig20},   where end of the study represents the censoring time.
The issues discussed in this section are however completely general to longitudinal study with staggered patient entry and not restricted to study with a specific (survival) outcome.
The key issue is that such  study designs are likely to suffer from potential ageing of the sample material,  prior to spectrometric measurement, such that sample age and duration to observation of the outcome  become confounded.
This problem may severely complicate future longitudinal studies. This point does not always seem to be realized in the present literature.
 This is probably also part due to the fact that genomic or gene expression data, as discussed before,  is likely not to be  affected by the same problem as would be the case for the more modern spectrometry-based omics data, as in proteomics for example.
  An excellent text describing modern survival analysis methodology in novel high-dimensional data applications was recently provided by van Houwelingen and Putter \cite{Houwelingen}[Part IV - Chapters 11 and 12].

It is at time of writing not clear how this issue should be addressed in design and analysis of longitudinal and survival studies with spectrometry-based omic data generally.
The problem is particularly important because it could affect all existing biobanks.
One could propose that instead of - or in addition to - the development of {\it biobanks} which store sample materials,  attention should be given to establishing {\it databanks} of (omic) spectra, which are measured at pre-specified and regular time points instead. Such an approach could break the confounding between the follow-up time of patients which is then de-coupled from the measurement times.
The problem with the latter proposal is that the measurement devices (spectrometers) themselves may exhibit ageing, such that the ageing problem is replaced with a spectrometer calibration problem.

\section{Discussion}
We have critically discussed  and contrasted the distinct issues of transformation, normalization and management of batch effect in the analysis of omic data.
Transformation, standardization and normalization are typically dealt with `after' the experiment and usually by the data analyst or statistician involved.
Batch effect and the presence thereof is an issue that should be considered both `before' and  `after' the data-generating experiment. The objectives here should be to optimize the design for known batch effect such that these cannot unduly affect conclusions or invalidate the experiment. This task is usually carried out in collaboration and prior discussion between both statistician and spectrometrist when planning the experiment.  Secondly,  appropriate methods should be used after the experiment to either eliminate or otherwise accommodate the batch effect after the experiment when analyzing the data.  The latter task will usually be carried out by the statistician solely.  Discussion of such methods falls outside the scope of this chapter.

Methodological choice in pre-processing of spectrometry data should in practice depend on many aspects.  Purpose of the study is a key consideration.  If prediction or diagnostic models are to be calibrated,  then pre-processing methods which can be applied `within-sample' without borrowing of information across distinct samples are more attractive because this can make subsequent calibration of any predictive rule easier.  Other considerations may be ease of communication of results,  established practice (in sofar it is reasonable of course), variance stabilization, interpretability and so forth. Our discussion has not been comprehensive but highlighted the main ideas and approaches instead,  pointing to common pitfalls, opportunities and future problems left to be solved.\\
\mbox{}
\\

\noindent {\bf Acknowledgements}\\
This work was supported by funding from the European Community's Seventh Framework Programme
FP7/2011: Marie Curie Initial Training Network MEDIASRES (''Novel Statistical Methodology for
Diagnostic/Prognostic and Therapeutic Studies and Systematic Reviews,'' www.mediasres-itn.eu) with the Grant
Agreement Number 290025 and by funding from the European Union's Seventh Framework Programme FP7/
Health/F5/2012: MIMOmics (''Methods for Integrated Analysis of Multiple Omics Datasets,'' http://www
.mimomics.eu) under the Grant Agreement Number 305280.
\\
\mbox{}
\\
Thanks to Mar Rodr\'iguez Girondo for critical comments on an early version of this text.

\bibliographystyle{agsm}
{}

\end{document}